# Shot noise in a strange metal

Liyang Chen[1], Dale T. Lowder[2], Emine Bakali[3], Aaron Maxwell Andrews[4], Werner Schrenk[5], Monika Waas[3], Robert Svagera[3], Gaku Eguchi[3], Lukas Prochaska[3], Yiming Wang[2], Chandan Setty[2], Shouvik Sur[2], Qimiao Si[2], Silke Paschen[3], Douglas Natelson[2,6,7*]

[1]Applied Physics Graduate Program, Rice University, 6100 Main St., Houston, TX 77005 USA

[2]Department of Physics and Astronomy, Rice Center for Quantum Materials, Rice University, 6100 Main St., Houston, TX 77005 USA

[3]Institute of Solid State Physics, TU Wien, Wiedner Hauptstraße 8-10, 1040 Vienna, Austria.

[4]Institute of Solid State Electronics, TU Wien, Gußhausstraße 25-25a, Gebäude CH, 1040 Vienna, Austria

[5]Center for Micro and Nanostructures, TU Wien, Gußhausstraße 25-25a, Gebäude CH, 1040 Vienna, Austria

[6]Department of Electrical and Computer Engineering, Rice University, 6100 Main St., Houston, TX 77005 USA

[7]Department of Materials Science and NanoEngineering, Rice University, 6100 Main St., Houston, TX 77005 USA

*Corresponding author: Douglas Natelson. Email: natelson@rice.edu

**Strange-metal behavior has been observed in materials ranging from high-temperature superconductors to heavy fermion metals. In conventional metals, current is carried by quasiparticles; although it has been suggested that quasiparticles are absent in strange metals, direct experimental evidence is lacking. We measured shot noise to probe the granularity of the current-carrying excitations in nanowires of the heavy fermion strange metal YbRh$_2$Si$_2$. When compared with conventional metals, shot noise in these nanowires is strongly suppressed. This suppression cannot be attributed to either electron-phonon or electron-electron interactions in a Fermi liquid, which suggests that the current is not carried by well-defined quasiparticles in the strange-metal regime that we probed. Our work sets the stage for similar studies of other strange metals.**

Strange metals are non-Fermi liquids that exhibit a low temperature ($T$) electrical resistivity contribution that is directly proportional to $T$. (*1*). This response that has been reported across many materials families, including cuprate (*2-4*) and pnictide (*5*) superconductors, ruthenates (*6*), heavy fermion metals (*7-9*), and twisted bilayer graphene (*10*). Strange metal properties typically arise at finite temperature above a quantum critical point (QCP), often in proximity to antiferromagnetic order (*11*). There are two broad classes of theories on metallic QCPs. Within the standard Landau approach of order parameter fluctuations, quasiparticles retain their integrity (*12, 13*). In approaches beyond the Landau framework (*14-18*), by contrast, no long-lived quasiparticles are expected to remain. Thus, determining the nature of the low energy current-carrying excitations is an important means to elucidate the nature of strange metals near QCPs.

How can we determine whether the current carriers in strange metals are quasiparticles? Shot noise in electrical conduction (*19*) is a unique probe of mesoscopic systems in which the current noise, $S_I = \langle (I - \langle I \rangle)^2 \rangle$, in a system driven out of equilibrium accesses the nature of the charge-carrying excitations. Here $I$ is the instantaneous current and $\langle I \rangle$ is the average current. The Fano factor, $F$, gives the ratio between the measured noise $S_I$ and $2e\langle I \rangle$, the expectation for Poissonian transport of ordinary "granular" charge carriers of magnitude $e$ with an average current $\langle I \rangle$. Shot noise has revealed fractionalization of charge in the fractional quantum Hall liquid (*20, 21*), fractional effective charges in quantum dot Kondo systems (*22, 23*), and pairing in superconducting nanostructures in the normal state (*24, 25*). A lack of granular quasiparticles would naively be expected to suppress shot noise, because the flow of a continuous fluid should have no fluctuations.

Despite their ubiquity, strange metals have yet to be examined through shot noise measurements for several technical reasons, and only few relevant theoretical predictions exist for any quantum critical systems (*26, 27*). In many materials strange metallicity is cut off at low temperatures by the onset of superconductivity, which complicates matters because shot noise measurements also require an electrical bias $eV$, where $e$ is the charge of the electron and $V$ is the applied voltage, large compared to the thermal scale $k_B T$ to distinguish from thermal noise, where $k_B$ is the Boltzmann constant. Tunneling transport into a strange metal faces the challenge that only discrete, individual electrons can be added or removed, likely leading to noise dominated by single-electron effects. Fortunately shot noise can be measured within a material using a diffusive mesoscopic wire, which requires nanofabrication of such structures without affecting electronic properties, a major challenge for many materials.

We have successfully made mesoscopic wires for noise measurements from epitaxial films of the heavy fermion material YbRh$_2$Si$_2$, a particularly well-defined strange metal (*9, 28*). YbRh$_2$Si$_2$ has a zero-temperature field-induced continuous quantum phase transition from a low-field antiferromagnetic heavy Fermi liquid metal to a paramagnetic one. The Hall effect displays a rapid isothermal crossover that extrapolates to a jump at the QCP in the zero-temperature limit, providing evidence for a sudden reconstruction of the Fermi surface across the QCP and an associated change in the nature of the quasiparticles between the two phases (*29*), as expected in the Kondo destruction description (*14-16*) for a beyond-Landau QCP. At finite temperatures, a quantum critical fan of strange metallicity extends over a broad range of temperature and magnetic field (*28, 30*). Recent time-domain THz transmission measurements (*31*) of the optical conductivity of epitaxial films of YbRh$_2$Si$_2$ reveal the presence of quantum critical charge fluctuations below 15 K, supporting the Kondo destruction picture in this system.

Measuring shot noise in YbRh$_2$Si$_2$ wires directly examines how current flows in a system thought to lack discrete charge excitations; these results can then be compared to predictions in Fermi liquids. We report measurements of shot noise in mesoscopic wires patterned from epitaxial films of YbRh$_2$Si$_2$, examined below 10 K, in the strange metal regime where phonon scattering is not expected to be relevant to the conductivity. The measured shot noise is found to be far smaller than both weak- and strong electron-electron scattering expectations for Fermi liquids, and also smaller than the values measured on a gold nanowire for comparison. Furthermore, the electron-phonon coupling determined experimentally using long YbRh$_2$Si$_2$ nanowires rules out strong electron-phonon scattering as a noise suppression mechanism. Therefore, the suppressed shot noise is evidence that current-carrying excitations in this strange metal defy a quasiparticle description in the examined temperature range. The no-quasiparticle model of Refs. (26, 27), despite being derived using conformal field theory for different kinds of QCP and associated phases than those of YbRh$_2$Si$_2$, predicts nontrivial bias- and temperature-dependent noise which is qualitatively consistent with the observed trends.

**Measuring shot noise in YbRh$_2$Si$_2$ devices**

High quality epitaxial films of YbRh$_2$Si$_2$ were grown by molecular beam epitaxy on germanium substrates (*31, 32*) (see Sect. 3 in (*33*) for details). The temperature dependence of the resistivity of these films above 3 K shows strange metal properties ($\rho = \rho_0 + AT^\alpha$, where $\alpha \approx 1$ in the low temperature limit, $\rho$ is the resistivity and $A$ is the temperature coefficient) as in the bulk material (Fig. 1C).

The films are patterned into nanowires through a combination of electron beam lithography and reactive ion etching (see Fig. S2 and accompanying discussion). The nanowire shown in Fig. 1B is 60 nm thick, 660 nm long, and 240 nm in width. Thick source and drain contact pads ensure that the dominant voltage measured under bias is across the nanowire, and act as thermal sinks(*34*). An important concern in fabricating nanostructures from strongly correlated materials is that the patterning process does not alter the underlying physics. As shown in Fig. 1C, the resistance $R(T)$ of the nanowire closely matches that of the unpatterned film, including a dominant linear-in-$T$ dependence at low temperatures. Similarly, in Fig. 1D the magnetoresistance (field in-plane, perpendicular to the current) in the nanowire is nearly identical to that of the unpatterned film, showing that the fabrication process did not alter the material's properties. This consistency also shows that the total $R$ is dominated by the wire, as the large contacts are coated in thick gold and would not exhibit such a magnetoresistance. Three nanowires patterned from this same film all show essentially identical transport and noise properties (data from devices #2 and #3 are shown in Fig. S5 in (*33*)).

The noise measurement technique is well-developed (*21, 24, 34*). A current bias is applied to the device by means of a heavily filtered voltage source and ballast resistors. Using a custom probe, the voltage across the device is measured through two parallel sets of amplifiers and a high-speed data acquisition system (Fig. S1C in (*33*)). The time-series data are cross-correlated and Fourier transformed to yield the voltage noise $S_V$ across the device, with the correlation mitigating the amplifier input noise (see Sect. 1 and 2 of (*33*) for a detailed discussion of calibration and averaging). Figure 2A shows the variation of the differential resistance, $dI/dV$, as a function of bias current, whereas Fig. 2B gives examples of voltage noise spectra at several bias currents at a base temperature of 3 K. At the maximum bias currents applied, the voltage drop across the wire is several mV, a bias energy scale considerably exceeding $k_B T$ (0.25 meV at 3 K), as needed for shot noise measurements.

**Theoretical expectations for the shot noise and Fano factor**

To understand the measured noise in YbRh$_2$Si$_2$, we first consider the expected current shot noise result for a diffusive metallic constriction. This is a long-established calculation within the Landauer-Büttiker formalism (*35-39*) for a Fermi gas (i.e., without any electron-electron interactions). A metal with well-defined quasiparticles is assumed in the source and drain, which obey the Fermi-Dirac (FD) distribution with a temperature set by the contacts, $T_0$. In the non-interacting, nanoscale limit, conduction takes place through spin-degenerate quantum channels with various transmittances, $\tau_i$. Each channel contributes to $S_I$ by an amount proportional to $\tau_i(1 - \tau_i)$. By averaging over the distribution of transmittances (*35-38*), one finds a predicted Fano factor $F \equiv S_I/2e\langle I \rangle = 1/3$. When inelastic electron-electron scattering is added to the otherwise non-interacting Fermi system, such that the system size along the direction of the current exceeds the electron-electron scattering length, $L > L_{ee}$ (see Fig. S11 in (*33*)) but is smaller than the electron-phonon scattering length, $L_{ph}$, there is a redistribution of energy and effective thermalization among the carriers (*40, 41*). There is a local quasi-thermal FD distribution within the wire described by a local electronic temperature $T_e(x)$ elevated above the lattice temperature, $T_0 = T$, assumed to be uniform and equal to the temperature of the contacts. This approach leads to a prediction of $F = \frac{\sqrt{3}}{4} \approx 0.433$ (*40, 41*). Fano factor predictions in both $L < L_{ee} < L_{ph}$ and $L_{ph} > L > L_{ee}$ limits have been confirmed in experiments in mesoscopic metal wires (*34, 42-44*).

In the present context, an important question is what happens in a Fermi liquid state when the electron-electron interactions are so strong that the quasiparticle weight is orders of magnitude smaller than the noninteracting case (= 1) for a free electron) and the Landau parameters are correspondingly large. It was recently shown that charge conservation constrains the Fano factor to be independent of the quasiparticle weight, and the combination of instantaneous electronic interactions and Poissonian charge transport dictates that the shot noise and average current get renormalized identically by the Landau parameters (*45*). As a result, for this regime (see solid line in Fig. S11 of (*33*)) that pertains to a strongly correlated Fermi liquid of interest here, the Fano factor would be $F = \frac{\sqrt{3}}{4} \approx 0.433$. (For further details see Sect. 13 of (*33*) and Ref. (*45*)).

Within the Fermi liquid quasiparticle picture, the only way to suppress shot noise below these levels is through strong electron-phonon scattering, which perturbs the electronic distribution function. In the limit of very strong electron-phonon coupling, the electronic distribution is constrained to be in equilibrium with the lattice temperature, $T_0$, and only Johnson-Nyquist noise at $T_0$ remains.

**Comparison to theoretical expectations**

Figure 3 shows the measured voltage noise as a function of bias current for a YbRh$_2$Si$_2$ nanowire, and its counterpart for a gold nanowire, for comparison. Shown as gray dot-dashed lines are the $F = \frac{1}{3}$ expectations, based on the measured differential resistance, $\left(\frac{dV}{dI}\right)$, as a function of bias current. Independent of any detailed analysis, the measured noise in the YbRh$_2$Si$_2$ wire is clearly suppressed well below the Fermi liquid expectation at all temperatures. Additional data on a two more wires (devices #2 and #3) are essentially identical (Sect. 7 and Fig. S5 of (*33*)). In contrast, the gold nanowire data (discussed further in Sect. 9 and Fig. S7 of (*33*)) are consistent with Fermi liquid predictions, with a slight suppression of the noise above 10 K as electron-phonon scattering becomes relevant (Fig. S7D of (*33*)).

The electron-phonon coupling may be extracted experimentally by analyzing the noise as a function of bias in a wire sufficiently long that electron-phonon scattering is dominant (*42*). As detailed in Sect. 5 of (*33*), using a 30 µm long YbRh$_2$Si$_2$ wire, we perform this analysis to determine the effective electron-phonon coupling in this material, finding a value sufficiently small that strong electron-phonon scattering is ruled out as a mechanism for suppressing the noise in the much shorter YbRh$_2$Si$_2$ nanowire constrictions.

Extracting effective Fano factors from the measured noise requires analysis in terms of finite temperature expressions for the shot noise. Subtleties about thermal noise can arise when the device is non-Ohmic, as discussed in Sect. 10 of (*33*), but corrections from the Ohmic case are small for the measured nonlinearities in Fig. 2A. The expected form for the current shot noise in an Ohmic system with Fano factor $F$ and differential resistance $\left(\frac{dV}{dI}\right)$ is (*19*)

$$S_I = F \cdot 2e\langle I \rangle \coth\left(\frac{eV}{2k_BT}\right) + (1-F)4k_BT\left(\frac{dV}{dI}\right)^{-1}. \qquad (1)$$

This expression reduces to the Johnson-Nyquist current noise $S_{I,JN} = 4k_BT\left(\frac{dV}{dI}\right)^{-1}_{I=0}$ in the zero bias limit and becomes $S_I = F \cdot 2e\langle I \rangle$ as expected in the high bias limit $eV \gg k_BT$. In the experiment we measure voltage noise, and for ease of comparison we subtract off the zero-bias

Johnson-Nyquist noise, so that effective Fano factors may be estimated by fitting to the voltage-based expression for the shot noise:

$$S_V = \left(\frac{dV}{dI}\right)_I^2 \left[F \cdot 2e\langle I\rangle \coth\left(\frac{eV}{2k_BT}\right) + (1-F)4k_BT\left(\frac{dV}{dI}\right)_I^{-1}\right] - 4k_BT\left(\frac{dV}{dI}\right)_{I=0}. \quad (2)$$

The fitted Fano factors of an YbRh$_2$Si$_2$ device and a gold nano wire device are shown in Fig. 4, which provides direct comparison between YbRh$_2$Si$_2$ and a Fermi liquid diffusive wire. Corrections stemming from non-Ohmic response lead to lower inferred Fano factors (Fig. S8 in (*33*)).

Our detailed thermal modeling of the present system under the standard Fermi liquid assumptions (see Sect. 6 of (*33*)) confirms that, including electronic thermal transport through the Wiedemann-Franz relation and the measured electron-phonon coupling, $F = \sqrt{3}/4$ would be expected in the present high bias limit. This is in sharp contrast to the experimental data of Fig. 3A. Experiments on bulk YbRh$_2$Si$_2$ crystals do not show large deviations from the Wiedemann-Franz relation in this temperature range (*46, 47*). Electronic transport measurements and thermodynamic measurements of YbRh$_2$Si$_2$ in this temperature regime, as well as THz optical conductivity measurements(*31*) in these films, show that phonons are not contributing strongly to the electronic properties in YbRh$_2$Si$_2$ below 15 K. As shown in Sect. 6 of (*33*) and Fig. S4, the measured electron-phonon coupling in YbRh$_2$Si$_2$ is too small by more than a factor of 35 to be responsible for the observed noise suppression.

To interpret these results, it is important to consider the nature of quasiparticles in terms of the single-particle spectral function and distribution functions. For a Fermi gas, the single-particle spectral function $A(k,\epsilon)$ at a given wavevector $k$ is a delta function in energy $\epsilon$ at $\epsilon = E_k$, where $E_k$ is the quasiparticle energy as a function of $k$, meaning that a particle excitation at $(k, E_k)$ in the zero temperature limit is perfectly well-defined in energy and has infinite lifetime with a spectral weight $Z = 1$. Correspondingly, the particle excitations follow the Fermi-Dirac distribution, and the Fermi surface is a perfectly sharp boundary at $T = 0$. In a Fermi liquid, the spectral function retains a peak for $k$ near the Fermi surface, which describes a quasiparticle with a nonzero spectral weight $Z < 1$. The distribution function near the Fermi surface is smeared but still has a non-zero discontinuity at $T = 0$ (*48*).

In the case of the particular type of non-Fermi liquid with a complete destruction of quasiparticles, one has $Z = 0$ everywhere on the Fermi surface. With such a complete smearing of the Fermi surface, there is no discontinuity in the distribution function even at $T = 0$. In this limit, when driven by a bias that does not greatly perturb the non-FD distribution function, there are no granular quasiparticles that carry the electrical current. We can then expect a much-reduced shot noise, as we observe in the form of a Fano factor that is considerably smaller than not only the strong electron-electron scattering expectation $F = \sqrt{3}/4$ but even the weak electron-electron scattering counterpart $F = 1/3$. We highlight this contrast in Fig. 4 and discuss it further in Sect. 13 (*33*). For reference, in the extreme case when the electron spectral function is entirely featureless, the continuous electron fluid would have no shot noise at all ($F = 0$). Interestingly, one approach to a quantum critical system with no quasiparticles (*26, 27*) predicts nonzero noise with a trend in bias and temperature that is quantitatively similar to that shown in Fig. 3A (as seen in Fig. S10 in (*33*)), though that model is based on a different form of quantum criticality (a superconductor-insulator transition) than that in YbRh$_2$Si$_2$. This is discussed further in Section 12 of (*33*).

**Discussion and outlook**

Shot noise is a probe that gives unique access to the nature of charge carriers. The suppressed noise shown in Fig. 3A and summarized in Fig. 4 is evidence that current in this strange metal regime is not governed by the transport of individual, granular quasiparticles. A Fano factor of zero is expected only for the most extreme case of a non-Fermi liquid that has a completely flat spectral function. A non-Fermi liquid that still has residual dispersive spectral features, despite a vanishing quasiparticle weight $Z$, would lead to a nonzero Fano factor. Any residual dispersive spectral features are naturally expected to somewhat sharpen as $T \to 0$, leading to a rise in $F$ as temperature is lowered, but in that case would never reach the $F = \sqrt{3}/4$ expectation for a strongly correlated Fermi liquid (where $Z$ is finite). The present experiment takes place firmly in the non-Fermi liquid regime (down nearly a factor of ten below the effective single-ion Kondo scale) already seen to exhibit critical scaling of the optical conductivity (*31*). As discussed further in Sect. 13 of (*33*), an effective Fermi liquid finite temperature correction to the expected Fano factor is not a naturally self-consistent explanation for the trend in Fig. 4. Even so, the present data are limited to 3 K and above; it is desirable to extend our measurements to below 3 K, which would allow for a direct comparison with the $T = 0$ theoretical expectations.

Although scattering techniques show incoherent, non-quasiparticle electronic response as a diffuse continuum across $(k, \epsilon)$, shot noise specifically targets the current-carrying excitations. The shot noise probes both the equilibrium non-Fermi liquid distribution function and its nonequilibrium evolution when perturbed by the difference in source and drain chemical potentials. In Fermi liquids, this approach has given insights into inelastic electron-electron scattering and the evolution of the nonequilibrium distribution function. This technique will provide crucial experimental constraints on such processes in materials in the strange metal regime. Moreover, strange metallicity as inferred from the resistivity is observed across many systems with quite disparate underlying microscopic physics (*2-8, 10, 28*). Shot noise provides an opportunity to test the extent to which these phenomenologically similar strange metals can fit within a single paradigm. We expect our work to trigger extensive further theoretical studies.

**Acknowledgments:**

The authors acknowledge helpful conversations with Matthew Foster and Andrew Lucas. Some of the noise measurement hardware was acquired through National Science Foundation award DMR-1704264.

**Funding:**

US Department of Energy, Basic Energy Sciences, Experimental Condensed Matter Physics award DE-FG02-06ER46337 (DN, LC, DTL).

US National Science Foundation DMR-1704264 (DN)

European Research Council ERC Advanced Grant 101055088 (GE, SP)

Austrian Science Fund FWF I4047 (EB, SP)

Austrian Research Promotion Agency FFG 2156529 (SP)

Austrian Science Fund FWF SFB F 86 (GE, SP); For the purpose of open access, the author has applied a CC BY public copyright licence to any Author Accepted Manuscript version arising from this submission.

European Union Horizon 2020 grant agreement 824109-EMP (EB, GE, LP, MW, RS, SP)

Austrian Research Promotion Agency FFG 883941 (WS, AMA)

US Air Force Office of Scientific Research FA8665-22-1-7170 (WS, AMA)

US National Science Foundation DMR-2220603 (YW, CS, SS, QS)

Robert A. Welch Foundation Grant C-1411 (YW, CS, SS, QS)

**Author contributions:**

Growth and characterization of the YbRh$_2$Si$_2$ films: EB, WS, MW, RS, GE, LP, AMA, SP

Patterning, device fabrication, noise measurements: LC, DTL

Noise data analysis and interpretation: LC, DN, QS, SP

Theoretical analysis of Fermi liquid effects: YW, CS, SS, QS

Critical insights on heavy fermion strange metals: QS, SP

Writing – initial draft: LC, DN, QS, SP

Writing – editing: LC, DTL, EB, WS, MW, RS, GE, LP, AMA, YW, CS, SS, QS, SP

**Competing interests:** The authors declare no competing financial interests.

**Data and materials availability:** All data presented in this paper and the supplementary material and the relevant analysis code are available through Zenodo (*49*).

**Supplemental Materials:**

Materials and methods

Supplementary Text

Figures S1-S11

References (50)-(52)

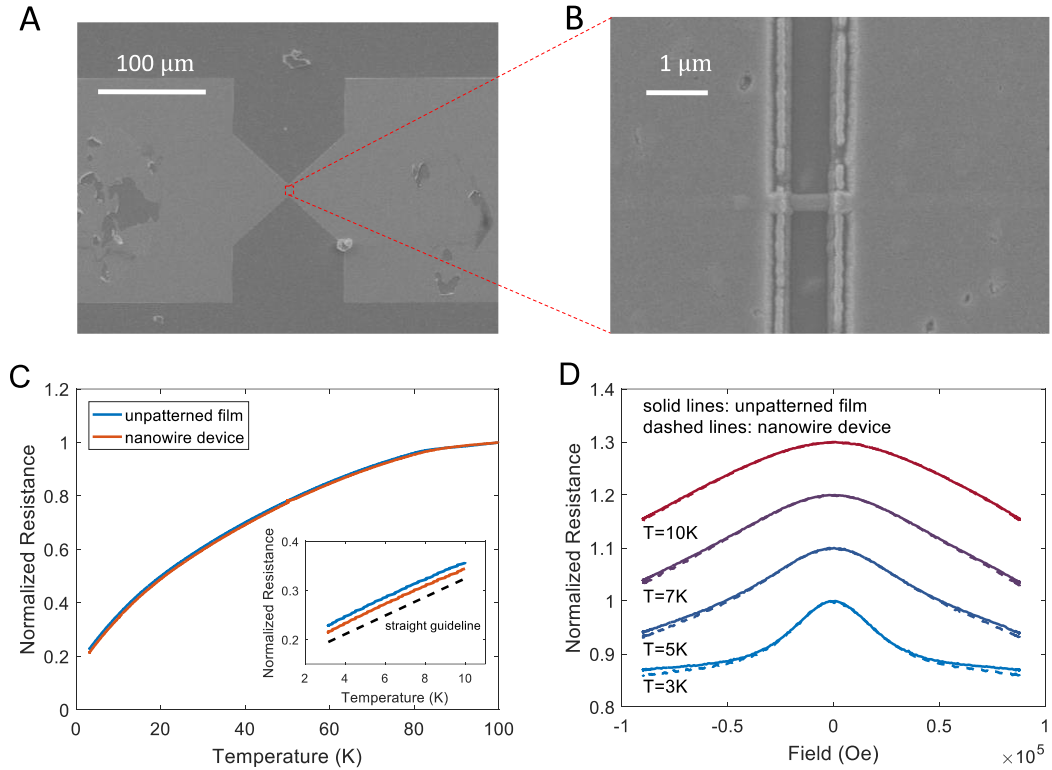

**Fig. 1. YbRh$_2$Si$_2$ nanowire device preparation and characterization.** (**A**) YbRh$_2$Si$_2$ nanowire between two large area, thick sputtered Au contacts on top of the unpatterned YbRh$_2$Si$_2$ film, deposited to ensure that the measured voltage is dominated by the nanowire. (**B**) Higher magnification view. Sample fabrication is discussed in detail in (*33*). (**C**) Normalized resistance as a function of temperature for both the unpatterned film and the etched nanowire, showing linear-in-*T* resistivity in the low temperature limit (inset), as seen previously (*31*). Unpatterned film resistance at 100 K = 17.8 Ω. Nanowire resistance at 100 K = 164.7 Ω. (**D**) Normalized resistance as a function of in-plane magnetic field for both the unpatterned MBE film and the etched nanowire (*B* oriented transverse to the nanowire), with curves shifted vertically for clarity. Zero field resistances for the film at 10 K, 7 K, 5 K, and 3 K are respectively 6.5 Ω, 5.5 Ω, 4.8 Ω, and 4.1 Ω. Zero field resistances for the wire at 10 K, 7 K, 5 K, and 3 K are 57.8 Ω, 49.0 Ω, 42.2 Ω, and 35.5 Ω. The nearly identical response between nanowire and unpatterned film confirms that patterning did not substantially alter the electronic properties of the epitaxial YbRh$_2$Si$_2$ material, and that the resistance is dominated by the wire.

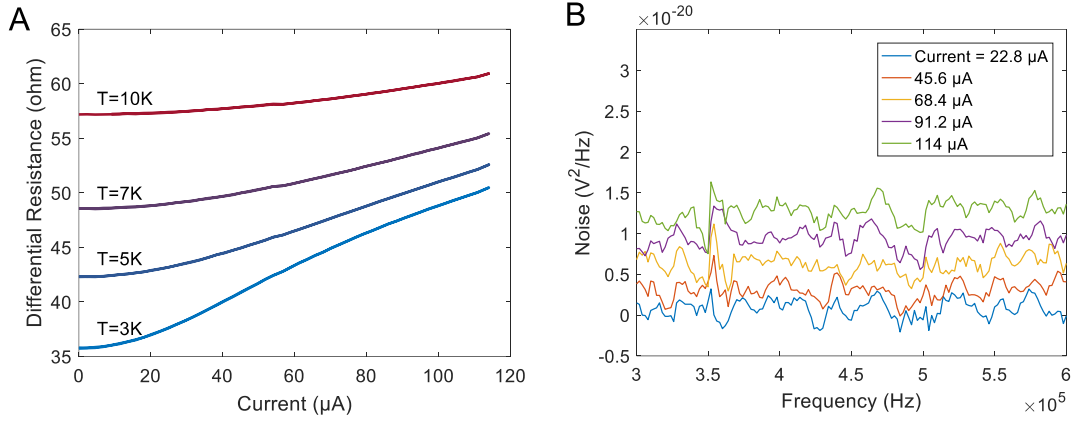

**Fig. 2**: **Noise characterization of a YbRh$_2$Si$_2$ nanowire.** (**A**) Differential resistance $dV/dI$ as a function of bias current at 10 K, 7 K, 5 K, and 3 K (top to bottom). Comparison with theoretical shot noise expectations requires this information (see Eqs. 1 and 2). (**B**) Averaged voltage noise spectra (with zero-bias spectra subtracted) of a YbRh$_2$Si$_2$ nanowire device at different bias levels at $T = 3$ K, over a bandwidth between 300 kHz and 600 kHz. This spectral range is free of extrinsic features and these voltage noise spectra are analyzed (Eq. 2) to determine the shot noise at each bias. Each spectrum shown is an average of 4500 spectra with 10 kHz bandwidth.

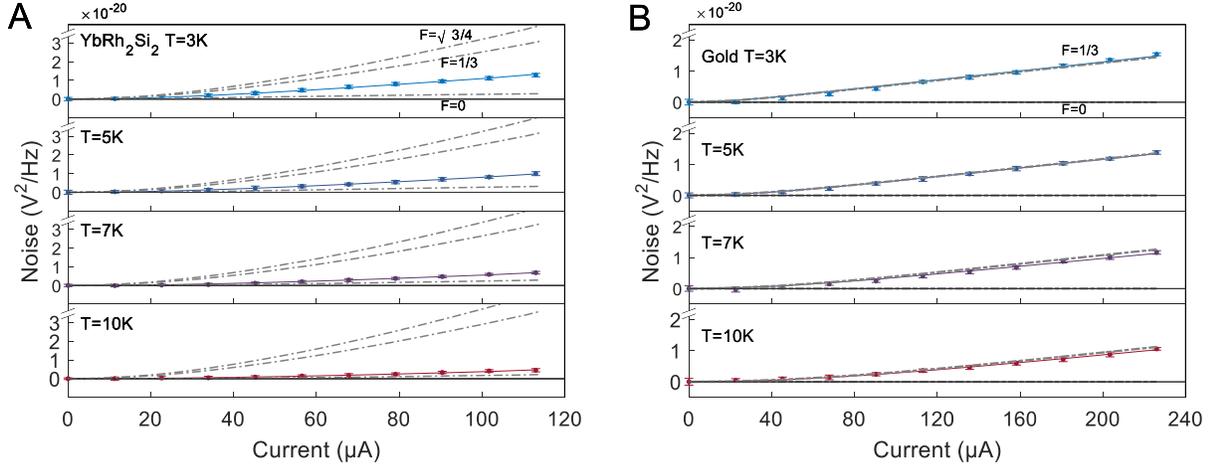

**Fig. 3. Noise vs. bias current characteristics.** (A) Noise vs. bias current for a YbRh$_2$Si$_2$ wire at various temperatures, with fits to Eq. 1 to extract effective Fano factors, for temperatures 10 K, 7 K, 5 K, and 3 K from bottom to top. Error bars are the standard deviation from 15 repeated bias-sweep measurements. Also shown for illustrative purposes are expectations for $F = \frac{\sqrt{3}}{4}, \frac{1}{3}$ and 0 (gray dot-dashed curves top to bottom, respectively) calculated using the measured differential resistance at each temperature, using Eq. 2. At all temperatures, the measured voltage noise is far below the theoretical expectations for shot noise in a diffusive nanowire of a Fermi liquid even in the weak electron-electron scattering limit. (B) Analogous data for a gold wire over the same temperature range, as discussed in Sect. 9 of (*33*). Data here are much closer to the $F = \frac{1}{3}$ Fermi liquid expectation, with the small deviation at the highest temperatures being attributed to electron-phonon scattering effects. Error bars are the standard deviation from 15 repeated bias-sweep measurements.

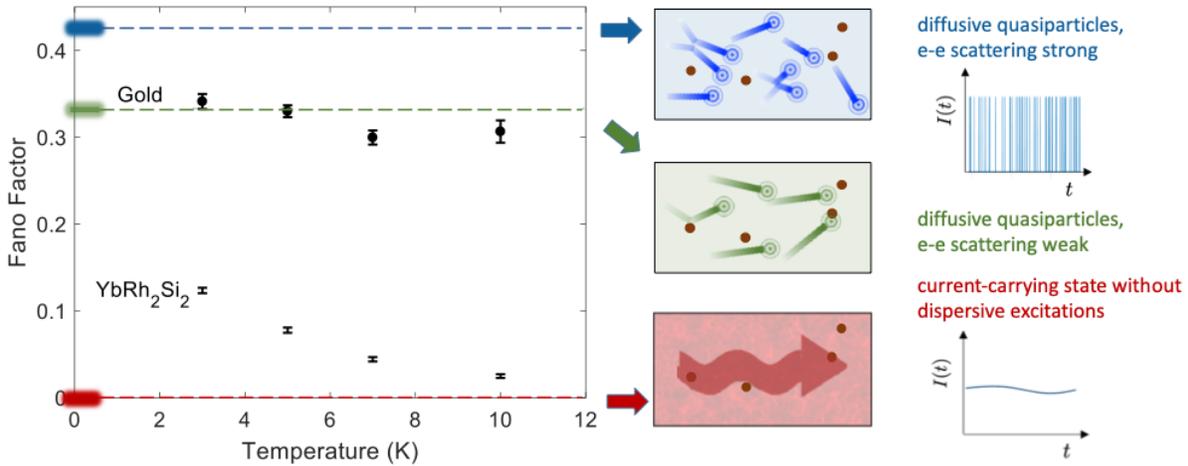

**Fig. 4**. **Fano factors and context for their interpretation.** The main panel shows Fano factors found from fitting the data in Fig. 3. Error bars are the standard error from fitting 15 repeated bias sweep measurements. In a Fermi liquid, current is carried by individual quasiparticle excitations, and the current as a function of time fluctuates with the arrival of each discrete transmitted carrier. Carriers scatter diffusively through static disorder (brown dots). When electron-electron scattering is weak (sample length $L < L_{ee}$), the expected Fano factor is $F = 1/3$ (green mark), while electron-phonon coupling can suppress this at higher temperatures. When electron-electron scattering is strong ($L > L_{ee}$), the expected Fano factor is $F = \sqrt{3}/4$ (blue mark). In a system without well-defined quasiparticles, charge transport is more continuous, leading to suppressed current fluctuations; in the extreme limit that electronic excitations are entirely non-dispersive, the Fano factor is expected to vanish (red mark). Dashed lines are guides to the eye.

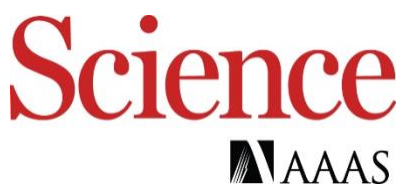

Supplementary Materials for

**Shot noise in a strange metal**

Liyang Chen[1], Dale T. Lowder[2], Emine Bakali[3], Aaron Maxwell Andrews[4], Werner Schrenk[5], Monika Waas[3], Robert Svagera[3], Gaku Eguchi[3], Lukas Prochaska[3], Yiming Wang[2], Chandan Setty[2], Shouvik Sur[2], Qimiao Si[2], Silke Paschen[3], Douglas Natelson[2,6,7*]

[1]Applied Physics Graduate Program, Rice University, 6100 Main St., Houston, TX 77005 USA

[2]Department of Physics and Astronomy, Rice Center for Quantum Materials, Rice University, 6100 Main St., Houston, TX 77005 USA

[3]Institute of Solid State Physics, TU Wien, Wiedner Hauptstraße 8-10, 1040 Vienna, Austria.

[4]Institute of Solid State Electronics, TU Wien, Gußhausstraße 25-25a, Gebäude CH, 1040 Vienna, Austria

[5]Center for Micro and Nanostructures, TU Wien, Gußhausstraße 25-25a, Gebäude CH, 1040 Vienna, Austria

[6]Department of Electrical and Computer Engineering, Rice University, 6100 Main St., Houston, TX 77005 USA

[7]Department of Materials Science and NanoEngineering, Rice University, 6100 Main St., Houston, TX 77005 USA

*Corresponding author: Douglas Natelson. Email: natelson@rice.edu

**This PDF file includes:**

Materials and Methods
Supplementary Text
Figs. S1 to S11
References (50)-(52)

**Materials and Methods**

S1. Noise measurement method

We used standard cross correlation methods to measure the noise spectra and corresponding intensity. The programmable voltage source (NI-DAQ6521) followed by two low pass *LC* filters (330 μF and 22 mH for the capacitance and inductance, respectively) provided a clean DC bias source, and two large current-limiting resistors (22.5 kΩ each) in series limit the current range. The samples were mounted on a customized low frequency measurement probe and loaded into a cryostat (Quantum Design PPMS). Two pairs of twisted wires shielded with stainless steel braided sleeving were used to apply bias, collect the noise signal, and reduce magnetic field induced noise. The sample, *LC* filters, transmission lines and first pair of pre-amplifiers are shielded by a Faraday cage to reduce environmental noise. The voltage noise generated from device is collected by two identical amplifiers chains, each consisting of two voltage preamplifiers (NF LI-75 and Stanford Research SR-560). Two amplified voltage signals are recorded by a high-speed oscilloscope (Picoscope 4262) at 2 MHz. Each spectrum is calculated using cross correlation of two voltage signals in 0.2 seconds containing 400,000 data points, and spectrum is averaged for 300 times (1 minute signals) in one measurement. We repeated measurements at each bias for 15 times to estimate the standard deviation of measured signal.

S2. Noise measurement setup calibration method

The room temperature thermal noise of a variety of resistors was used to calibrate the setup. The voltage noise power spectral density found from the cross-correlation can be expressed as

$$S_V(f) = 4k_B TR \times A$$

where $4k_B TR$ is the Johnson-Nyquist voltage noise at the resistor $R$ and $A$ is a coefficient containing the squared amplifier gain and a numerical factor related to the cross-correlation parameters (number of data points of each time series, the sampling frequency and the Hanning window for the Fourier transform).

The thermal noise spectra for different resistors are shown in Fig. S1A. The lowest spectrum is corresponding to thermal noise from a 50$\Omega$ resistor, and highest one is from a 391$\Omega$ resistor. The scattered dots are the raw 5 Hz bandwidth spectrum and lines are the averaged spectrum over 1 kHz bandwidth. The spectrum from 300 kHz to 600 kHz is almost flat, so we used the mean value of this spectrum as the voltage noise power.

To find the coefficient $A$, in Fig. S1B, we plotted the mean values of the raw spectra versus resistance $R$. The noise intensity for these resistors fall on the fitting straight line, showing that the zero-bias voltage noise power is strictly linearly proportional to the resistance, as expected for Johnson-Nyquist thermal noise. A linear regression (shown by the yellow line) finds the relation between voltage noise power and resistance:

$$S_V(f) = 4k_B TR \times A = 1.34 \times 10^{-11} \times R + 1.05 \times 10^{-11}\, V^2/Hz$$

The intercept comes from noise background of our measurement system, which brings a constant offset to noise spectra in a series of measurement, but it does not affect our results for the bias dependence of the noise. The fitting is shown in linear scale to see all data clearly. From the formula above, we find the value of $A = 1.34 \times 10^{-11}/4 k_B T = 8.23 \times 10^8$. From our definition, $A$ is dimensionless.

S3. <u>Materials: Film growth</u>

YbRh$_2$Si$_2$ thin films were grown on Ge (001) wafers in a RIBER C21 EB 200 molecular beam epitaxy (MBE) system, using a high-temperature (RIBER HT 12) cell for Rh, a medium high-temperature (MHT) cell for the Si, and a dual zone low-temperature (DZ-MM) cell for Yb. The temperature-dependent growth rates for Rh and Si were determined with elemental films, that of Yb with a flux gauge. With these calibrations, the cell temperatures were set for stoichiometric growth. A series of films was grown at different rates. The film selected for the present study was closest to the ideal 1:2:2 stoichiometry according to energy-dispersive X-ray spectroscopy (EDX) measurements, showed no foreign phase in x-ray diffraction, had a smooth surface in atomic force microscopy (AFM), and the highest residual resistance ratio (*32*). The quality is comparable to that of the films studied in Ref. (*31*), which were grown with electron beam evaporators instead of Knudsen cells for Rh and Si.

S4. <u>Materials: Reactive ion etch, device fabrication, and wire properties</u>

To make nanodevices from the film, we need to etch away part of film to define the device geometry. Because of the robustness of YbRh$_2$Si$_2$ film, it is difficult to find a proper chemical etchant or reactive ion etch (RIE) recipe to etch the YbRh$_2$Si$_2$ selectively without damaging the Ge substrate; most etchants remove the Ge faster than the YbRh$_2$Si$_2$. We choose to use pure argon plasma in a RIE system to etch the film through physical sputtering, which has a similar etch rate for both the YbRh$_2$Si$_2$ film and the substrate. The etch rate of the film under argon plasma depends on the radio frequency (RF) and inductively coupled plasma (ICP) power, but also heavily depends on the pressure. Lower pressure means lower scattering probability of argon atoms before hitting the film surface, and results in higher argon atom energy. We found that if the working pressure is comparatively high (such as 20 mtorr for usual RIE recipe), there will be many sharp islands left on the etched surface, most likely because the low energy argon atoms only destroy the surface structure, but do not effectively remove the YbRh$_2$Si$_2$ material. If the working pressure is too low, there is insufficient plasma density to etch film. After trials with different pressures, we found that the following recipe works well in our Oxford Plasmalab System 100/ICP 180 etcher: pressure 4 mtorr, argon gas flow rate 35 sccm, RF power 185 W (corresponding DC voltage 400 V), ICP power 700 W. The edge of a region of film following 1 minute RIE etching in these conditions is shown in an atomic force microscopy (AFM) in Fig. S2A with a line-cut topograph in Fig. S2B. PMMA 950 was used as the protective cover of the unetched region and removed by warm acetone with brief sonication, and low energy oxygen plasma. The AFM scan image shows the etch rate using this recipe is about 47.5 nm per minutes.

The devices are made by several steps of e-beam lithography, sputtering, and etching. The steps are summarized in Fig. S2C-F. The first row displays the view from top, and second row shows the cross-sectional view of the same steps. In first step, standard e-beam lithography on PMMA 950/495 double layer resist defines the patterns for the two source and drain gold contacts. 200 nm thickness gold contacts pads and a 60 nm Cr hard mask layer for RIE etch are deposited via sputtering, followed by liftoff. In the second step, another round of e-beam lithography and sputter deposition makes the Cr nanowire that serves as the etch mask for the nanowire device. Two minutes of argon RIE under the conditions described above etch away the exposed YbRh$_2$Si$_2$ film. Finally, the Cr masks are removed by soaking in 70°C 37% HCl solution, leaving the gold pads and protected YbRh$_2$Si$_2$ nano wires. Wire bonding is used to make connection from the device to our customized probe. A small misalignment in the first patterning step left the edges of the gold pads exposed to RIE, causing edge roughness seen in Fig. 1A, but this does not affect our measurements, because the gold pads close to the nanowires are protected well by the Cr nano wire mask in the second patterning step.

Any estimate of an effective carrier mean free path is necessarily crude and likely ill-posed, as the concept of quasiparticles is in doubt in the strange metal regime. Moreover, even in the Fermi liquid regime, YbRh$_2$Si$_2$ is a multiband conductor with strongly renormalized quasiparticles. For completeness, an order of magnitude estimate of an effective mean free path may be found from(47) $l \approx \sigma \frac{\hbar}{e^2}(3\pi^2)^{\frac{1}{3}} n^{\frac{-2}{3}}$, where $\sigma$ is the conductivity and $n$ is the carrier density, estimated either from the Hall effect ($2.6 \times 10^{28}$ m$^{-3}$, though the multiband conduction makes this rough) or superfluid density in the superconducting state ($4.86 \times 10^{27}$ m$^3$). In the unpatterned film, the 2 K conductivity is $1.3 \times 10^6$ S/m, giving an estimate of an effective mean free path of between 9 nm and 28 nm. The residual resistivity of the nanowires is higher than that of the unpatterned film due to edge scattering and static disorder introduced in the patterning/etching processes, so the effective mean free path would be even shorter.

**Supplementary text**

S5. Inferring the electron-phonon coupling in YbRh$_2$Si$_2$

In a Fermi liquid, strong electron-phonon scattering can suppress shot noise by anchoring the quasiparticle distribution function to the temperature of the lattice. As shown by Henny *et al.* in Ref. (*41*), the electron-phonon coupling strength $\Gamma$ can be determined experimentally by measuring the noise as a function of bias current in a wire much longer than the electron-phonon scattering lengthscale. In this limit, the electron temperature profile within the wire is modeled by the equation

$$\frac{\pi^2}{6}\frac{d^2 T_e^2}{dx^2} = -\left(\frac{eE}{k_B}\right)^2 + \Gamma\left(T_e^5 - T_{ph}^5\right) \quad , \tag{S1}$$

where $T_e$ is the local electron temperature, $x$ the position on the nanowire, $E$ the local electric field, and $T_{ph}$ the phonon temperature, which is assumed to be spatially uniform, equal to the base temperature, and is also the electronic temperature in the contacts. The model is derived assuming that all the Joule heating power within a small segment of wire is conducted away via electronic thermal conductivity or lost to the phonons. Accounting for the measured temperature dependence of the electrical conductivity and assuming the electrical conductivity only depends on local temperature complicates the expression slightly to

$$\frac{\pi^2}{6}\frac{d^2 T_e^2}{dx^2} - \frac{\pi^2 T_e}{3r}\frac{dr}{dT_e}\left(\frac{dT_e}{dx}\right)^2 = -\left(\frac{eE}{k_B}\right)^2 + \Gamma\left(T_e^5 - T_{ph}^5\right) \quad . \tag{S2}$$

The resulting model may be solved numerically to find a consistent $T_e(x)$ for a given bias current. Given this, the integrated thermal noise may be computed by adding up the contributions of each segment of wire: $S_V = \int 4k_B r(T_e(x)) T_e(x) dx$, where $r$ is the local resistance per unit length.

Figure S3A shows the measured voltage noise vs. bias current data at 3 K, 5 K, and 7 K for a 30 μm long YbRh$_2$Si$_2$ wire of comparable width to the short nanowires, fabricated through the same process. For a long nanowire with a length much longer than the electron-phonon scattering length, thermal transport is dominated by electron-phonon coupling. Following the method of Henny *et al.* in Ref. (*41*), we numerically solved the equation $\left(\frac{eE(x)}{k_B}\right)^2 = \Gamma\left(T_e(x)^5 - T_{ph}^5\right)$. At each temperature the data can be fit extremely well with a single $\Gamma$, the values of which are $9 \times 10^9$ K$^{-3}$m$^{-2}$, $9.5 \times 10^9$ K$^{-3}$m$^{-2}$, and $10 \times 10^9$ K$^{-3}$m$^{-2}$, for temperatures 3 K, 5 K, and 7 K respectively. Solving the full Eq. (S2) leads to slightly smaller (by ∼ 5%) inferred values for $\Gamma$. These values are of the same order as the reported coupling for gold at 2 K, $5 \times 10^9$ K$^{-3}$m$^{-2}$. As is shown below explicitly in Sect. 6, with these measured values for $\Gamma$, electron-phonon coupling cannot be responsible for the observed shot noise suppression.

S6. Electron heating effects and resulting noise

In a normal diffusive metal wire with length much longer than the electron-electron length and much shorter than the electron-phonon interaction length, electron-electron scattering is predicted to lead to electronic heating, with an elevated electron temperature profile along the wire, and cause a corresponding noise intensity increase, with a high bias limit corresponding Fano factor $F = \sqrt{3}/4$. Note that relaxing the boundary condition at the ends of the wire and allowing the pads to increase in temperature as well would lead to predictions of even greater noise.

Electron-phonon coupling effects can suppress the noise in the usual Fermi liquid quasiparticle scenario. YbRh$_2$Si$_2$ has a Debye temperature more than twice that of gold (*50*), which already makes that scenario unlikely in the present case. As explained here, we are able to *rule out* electron-phonon scattering as the mechanism responsible for the measured suppressed noise in the Yb Rh$_2$Si$_2$ nanowires. We consider the more sophisticated model based on Eq. (S2), a well-developed 1D model for the local temperature profile (*34, 41, 42*), and take two approaches.

First, we consider what the expected noise *would* be if there were strong electron-electron scattering (resulting in an effective $T_e(x)$) as well as electron-phonon scattering consistent with the experimentally determined value for $\Gamma$ in Yb$_2$Rh$_2$Si$_2$ from Sect. 5 and Fig. S3. We compute the temperature profile $T_e(x)$ expected within the wire at 3 K and $\langle I \rangle \approx 113 \mu A$, and then compute the expected noise from $S_V = \int 4 k_B r(T_e(x)) T_e(x) dx$ and the expected non-Ohmic response due to that $T_e(x)$. The results are shown in Fig. S4A-C. Using the measured $\Gamma$ in Yb$_2$Rh$_2$Si$_2$, the predicted noise and the predicted non-Ohmic response are much greater than what is actually seen in the experiments. The measured noise response is not compatible with the experimentally determined electron-phonon coupling.

In our second proof-by-contradiction approach (Fig. S4D), we work backward from the measured noise to estimate what electron-phonon coupling parameter would be necessary to match the experimental noise data at each current within the quasiparticle heating model. For the suppressed noise to result from electron-phonon scattering, $\Gamma$ in Yb$_2$Rh$_2$Si$_2$ would have to be $3.4 \times 10^{11}$ K$^{-3}$m$^{-2}$, which is about 35 times higher than the experimentally measured value. Again, we find that the measured electron-phonon coupling is incompatible with electron-phonon scattering as the mechanism responsible for the suppressed noise in the short nanowires.

S7. Data on additional short nanowire devices

We have performed measurements on two additional nanowires with virtually identical results. Figure S5 presents the additional data sets on devices #2 and #3 fabricated on the same chip as the one in the main text.

S8. Additional data at 9 T in-plane transverse magnetic field

While the differential resistance changes with increasing in-plane magnetic field, as shown in Fig. 1D, apart from this there is no significant change in the noise response. In Fig. S6 are data on device #2 taken in an in-plane field of 9 T. Zero-field data for comparison are in Fig. S5B, C.

S9. Data on a gold wire for comparison

We performed the analogous measurements on a gold nano wire to compare the YbRh$_2$Si$_2$ with a conventional metal. The gold nano wire is made in two steps. In the first step, 18 nm gold and 1 nm titanium adhesion layer are patterned by e-beam lithography and deposited by e-beam evaporator. In the next step, 200 nm thick gold pads are deposited by evaporation with a 2 nm Ti adhesion layer. The structure is shown in Fig. S7A. The voltage noise power values and Fano factors are much closer to the conventional Fermi liquid expectations, as shown in Fig. S7C,D. The slight decrease in Fano factor starting at 10 K and above is expected to be a consequence of electron-phonon scattering.

S10. Noise analysis for a non-Ohmic device

Experimentally, the noise in the YbRh$_2$Si$_2$ nanowires is unequivocally suppressed relative to Fermi liquid expectations. When attempting to extract an effective Fano factor, it is important to consider whether non-Ohmic response could be influencing the analysis. There are a number of approaches that try to address the issue of thermal noise in an intrinsically non-Ohmic device (e.g. (*51*), (*52*)). The situation is particularly challenging in a mesoscopic device driven out of equilibrium, where it is not generally possible to cleanly separate the measured noise into contributions that are purely thermal and purely shot noise. Note that if *all* the apparent non-Ohmic response in a device originates from local heating of the electrons, as in the Fermi liquid $F = \sqrt{3}/4$ situation discussed in Sect. 6, then Eq. 1 in the main text is the correct analysis. The concern arises if non-Ohmic response is intrinsic to the device.

For a passive device without large reactive contributions, the typical approach in assessing the thermal noise is to consider a term with a nonlinear correction:

$$S_{V,Th}(I) = 4k_B T \left( \left(\frac{dV}{dI}\right)_I + \frac{1}{2}\langle I \rangle \left(\frac{d^2 V}{dI^2}\right)_I \right)$$

If we take the conservative approach and assume that all of the observed nonlinearity in Fig. 2A is intrinsic and not related to local electron temperature changes, we can analyze the data of Fig. 3A with this approach, as shown in Fig. S8. In the absence of a nonlinearity correction, Fano factors can be found using Eq. (S3):

$$S_V = \left(\frac{dV}{dI}\right)_I^2 \left[ F \cdot 2e\langle I \rangle \coth\left(\frac{eV}{2k_B T}\right) + (1-F)4k_B T \left(\frac{dV}{dI}\right)_I^{-1} \right] - 4k_B T \left(\frac{dV}{dI}\right)_{I=0} \quad (S3)$$

and using the nonlinearity correction, Fano factors can be inferred using Eq. S4:

$$S_V = \left(\frac{dV}{dI}\right)_I^2 \left[F \cdot 2e\langle I\rangle \coth\left(\frac{eV}{2k_BT}\right) + (1-F)\left(4k_BT\left(\frac{dV}{dI}\right)_I^{-1} + 2k_BT\langle I\rangle\left(\frac{d^2V}{dI^2}\right)_I \Big/ \left(\frac{dV}{dI}\right)_I^2\right)\right] - 4k_BT\left(\frac{dV}{dI}\right)_{I=0}.  \tag{S4}$$

We note that in a general approach commonly used that doesn't explicitly consider non-Ohmic properties of devices, the current shot noise is be expressed as:

$$S_I = 4k_BTG + 2e\langle I\rangle F\left[\coth\left(\frac{eV}{2K_BT}\right) - \frac{2k_BT}{eV}\right] \tag{S5}$$

For comparison, we also use this formula to extract the Fano factor through replacing conductance $G$ with $\langle I\rangle/V$ and rewriting it in voltage noise format:

$$S_V = \left(\frac{dV}{dI}\right)^2\left[F \cdot 2e\langle I\rangle \coth\left(\frac{eV}{2K_BT}\right) + (1-F)\frac{4k_BT\langle I\rangle}{V}\right] - 4k_BT\left(\frac{dV}{dI}\right)_{I=0} \tag{S6}$$

The fitting curves for this approach are shown in Fig. S8C and the Fano factors obtained using this method are shown in Fig. S8 D.

The suppression of the measured noise below the Fermi liquid expectations is unambiguous, as pointed out in the main text, and the differences between the inferred Fano factors with and without the nonlinearity correction are small.

S11. Noise suppression from the distribution function perspective

Figure S9 gives a qualitative description of the noise based on electronic distribution functions (*34*). In the Fermi liquid case (*39*) with $Z$ close to 1, the quasiparticles in the source and drain are well described by Fermi-Dirac (FD) distributions with a substrate temperature $T_0$ shifted by the applied bias scale $eV$. At the midpoint of the wire, in the absence of electron-electron scattering, the distribution function would have a step, as shown, and the consequence of this for noise in the diffusive limit is $F = 1/3$. In the limit that inelastic scattering between electrons is fast compared to traversal of the wire, the energy is redistributed among the quasiparticles such that the distribution is approximately FD with an elevated electron temperature $T_e(x)$ evaluated at the wire midpoint, which leads to $F = \sqrt{3}/4$. In the case of a strange metal without quasiparticles, the situation is considerably more complicated, with an equilibrium distribution function that differs strongly from the FD case. To our knowledge, the nonequilibrium situation remains to be addressed.

S12. One theoretical model for current fluctuations in a quantum critical metal

One explicit discussion of nonequilibrium noise in a quantum critical metal exists in the literature (*26, 27*). It serves as a valuable example of a model without quasiparticles that nonetheless predicts nonzero shot noise as a function of bias. The applicability of the model in question, however, is limited and not necessarily well-suited to this experiment. That model considers an entirely different quantum critical metal, one associated with the Bose-Hubbard model of the superconductor-insulator transition. The approach is based on holography and maps between the nonequilibrium current fluctuations in the quantum critical metal and Hawking radiation within an effective black hole model. The result is an expression that interpolates between Johnson-Nyquist noise at zero bias and shot noise that scales like $\sqrt{V}$ at high bias, $S_I = 4k_B T^*/\left(\frac{dV}{dI}\right)$, where $\pi k_B T^* = [(\pi k_B T)^4 + \hbar^2 c_0^2 e^2 E^2]^{1/4}$, $c_0$ is a characteristic velocity, and $E = V/L$ is the driving electric field along the device. It is worth noting that this model as described has issues when extrapolated to the zero-temperature limit, as the predicted Fano factor diverges like $E^{-1/2}$ when approaching zero bias.

While this model is for an entirely different kind of quantum critical point, it is interesting for illustrative purposes to apply it to the measured data. Fitting the lowest temperature data best constrains $c_0$. We fit the 3 K noise as a function of bias current with this expression using $c_0$ as the only adjustable parameter. Comparing (Fig. S10) with the measured data at higher temperatures and no further adjustable parameters shows good agreement for $c_0 = 6.6 \times 10^5$ m/s. This may be fortuitous and a consequence of the limited bias range available. For the value of $c_0$ above, a bias range a factor of two larger or more would be needed to determine unambiguously whether the noise scales as $E^{1/2}$ in the high bias limit. For the present samples and temperatures this would raise strong concerns about self-heating. Extending measurements to much lower temperatures may allow testing of whether this high bias scaling holds in this system, despite the model's derivation for a different type of quantum critical system.

## S13. Fano factor in strongly correlated Fermi liquids

Shot noise in a diffusive metal is sensitive to the relative size of various length scales at play. Generically, four "intrinsic" length scales are present, viz. the electron-electron scattering length ($L_{ee} \sim \sqrt{D\tau_{ee}}$), electron-impurity scattering length ($l \sim \sqrt{D\tau_{imp}}$), electron-phonon scattering length ($L_{ph} \sim \sqrt{D\tau_{ph}}$), and the average electron density ($\rho_0^{-1/3} \sim k_F^{-1}$), where $D$ is the diffusion constant. In the regime of interest in YbRh$_2$Si$_2$, $L_{ph}$ ($k_F^{-1}$) is the largest (smallest) length scale. The impurities are sufficiently dilute such that $l \gg k_F^{-1}$ (29). In a strongly correlated Fermi liquid, $L_{ee}^{-1} \sim (T^2 + V^2)$, with $T$ ($V$) being the temperature set by the contacts (applied voltage). We consider the regime of $T$ and $V$ such that $L_{ee} \gg l$.

First, though, we summarize the known results that have been derived in a Fermi-gas-based approach. Depending on the linear dimension of the system, $L$, the Fano factor has three distinct behaviors, viz. (i) $F = 1/3$ for $l \ll L \ll L_{ee} \ll L_{ph}$; (ii) $F = \sqrt{3}/4$ for $l \ll L_{ee} \ll L \ll L_{ph}$; (iii) $F \sim L^{-\alpha}$ for $l \ll L_{ee} \ll L_{ph} \ll L$. These results are summarized in Fig. S11. The Fermi-gas-based approach studies a Fermi gas for regime (i) or for regime (ii), and considers the role of electron-electron scattering rate on the non-equilibrium distribution function but does not take into account the effects of either the Landau parameters or the correlation-induced reduction of the quasiparticle weight (from the noninteracting value) (40, 41).

In the present context, the question is what happens in a Fermi liquid where the quasiparticle weight $Z$ can be renormalized down by orders of magnitude and the Landau parameters can be correspondingly large. Regime (ii) pertains to such a strongly correlated Fermi liquid. In Ref. (45) some of us demonstrated that $F = \sqrt{3}/4$ even in this case. Through a microscopic derivation of the Boltzmann-Langevin transport equation for a correlated diffusive metal, it was shown that charge conservation constrains the Fano factor to be independent of the quasiparticle weight. Furthermore, the combination of instantaneous electronic interactions and Poissonian charge tunnelings dictate that the shot noise and average current get renormalized identically by the Landau parameters. As a result, strong correlations lead to a Fano factor of $F = \sqrt{3}/4$, regardless of how large the Landau parameters are. Thus, a strong deviation from this expected value for the Fano factor in a strongly correlated diffusive metal indicates a metallic state that goes beyond a quasiparticle description.

While the resistivity of the measured device is firmly in the strange metal regime, one can still consider the possible effect of finite-temperature Fermi liquid corrections to the Fano factor. Since the only temperature scale in this picture is $T_F$, and the leading temperature correction of a Fermi liquid is of the form $T^2$, we can expect that the Fano factor at nonzero temperatures, $F(T)$, is of the form:

$$F(T) = \left(\frac{\sqrt{3}}{4}\right)\left(1 - C\left(\frac{T}{T_F}\right)^2\right) \tag{S7}$$

where $C$ is a number of order unity.

For the present system, taking the effective Fermi temperature scale of YbRh$_2$Si$_2$ as the effective single-ion Kondo scale $\approx 25$ K, we infer that, at 3 K, the quantity $\left(\frac{T}{T_K}\right)^2 \approx 0.015$. This would imply a reduction on the order of 1.5% of the Fano factor, from ($\sqrt{3}/4$)=0.433 to about 0.427, which is about 3.4 times the measured value of about 0.125.

Conversely, if one assumed that the observed low Fano factor is due to such a Fermi liquid correction and attempted to infer an effective $T_F$ from Eq. (S7) and the trend in Fig. 4, a rough extrapolation would imply $T_F \sim 2$ K, below the experimentally measured regime. Thus, in the observed temperature range, such a finite temperature Fermi liquid correction cannot self-consistently explain the measured suppression of the Fano factor.

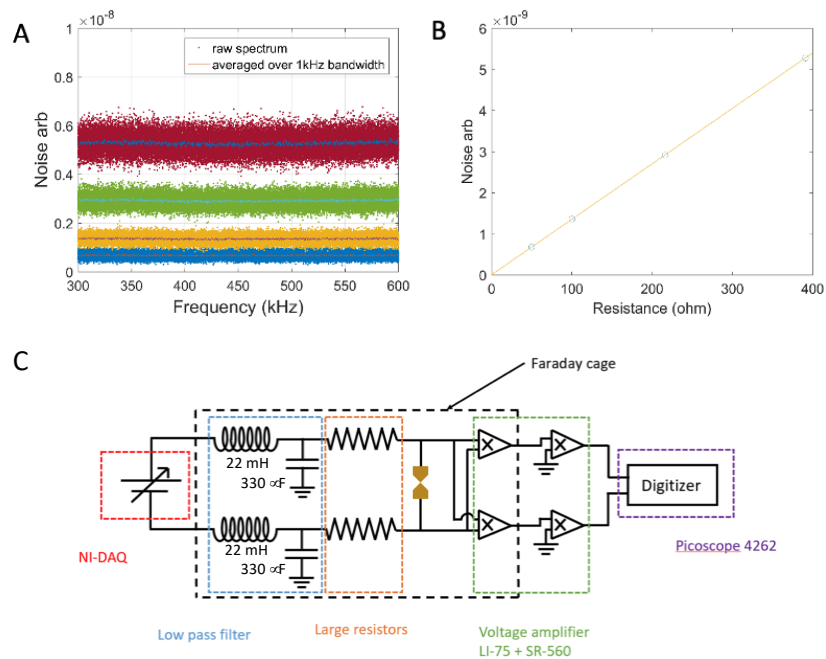

**Fig. S1. Noise measurement setup calibration using resistors' thermal noise.** (A) Thermal noise spectra for resistors equal 50 Ω, 100 Ω, 216 Ω, 391 Ω, at room temperature from 300 kHz to 600 kHz. The spectra increase linearly with resistance. The scattered dots represent the raw spectra with 5 Hz resolution bandwidth, and solid lines are for the spectra averaged over 1 kHz bandwidth. (B) Voltage noise power dependence on resistance values. Noise power values are calculated using the mean values of spectra in (A) and shown by blue circles. Linear fitting is used to calculate the effective gain of whole system, and we obtained $8.23 \times 10^8$ for our setup. (C) Diagram of measurement circuit, showing the amplifier chains in parallel across the device (gold) and the filters to eliminate noise from the biasing circuitry.

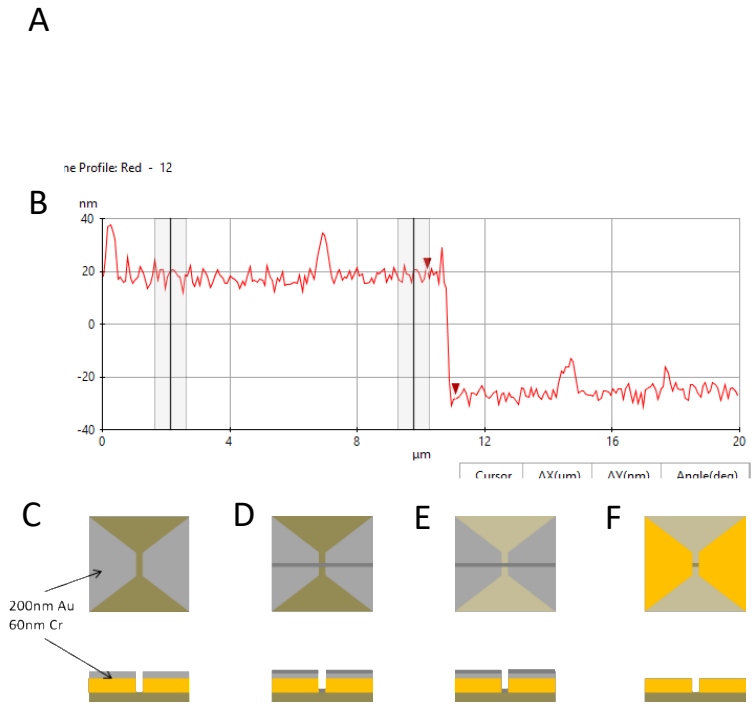

**Fig. S2. Device fabrication.** (A) Surface topograph scanned by AFM on edge of argon etched film in a 3 μm × 20 μm area. Red horizontal line indicates the position of the side view in (B), and red triangles are the two positions used to calculate etching rate. (B) Thickness change of etched film under one minute argon reactive ion etching. Two gray windows represent the two ranges on protected film to flatten the tilted surface and correct corresponding error. The AFM surface topograph shows the film was etched by about 47.5 nm in one minute. (C) Two 200 nm gold pads and 60 nm Cr hard mask on top are patterned by ebeam lithography and deposited using sputtering. Leaving small gap in middle for nano wires. The diagram on bottom shows the side view of same process. Film, gold pads and Cr masks are indicated by dark brown, yellow and light gray respectively. (D) Nanowire Cr mask is made using similar process as gold pads, shown by the dark gray line. The width of nano wires varies from 150 nm to 300 nm. (E) The uncovered part of the YbRh$_2$Si$_2$ film is etched in argon reactive ion etch for 2 minutes using the recipe in Sect. 4. The light brown color represents the etched part, leaving only Ge substrate. (F) The Cr hard masks for gold pads and nanowires are removed by warm (70°C) concentrated HCl solution (37%), leaving only golds contact pads and YbRh$_2$Si$_2$ nano wires.

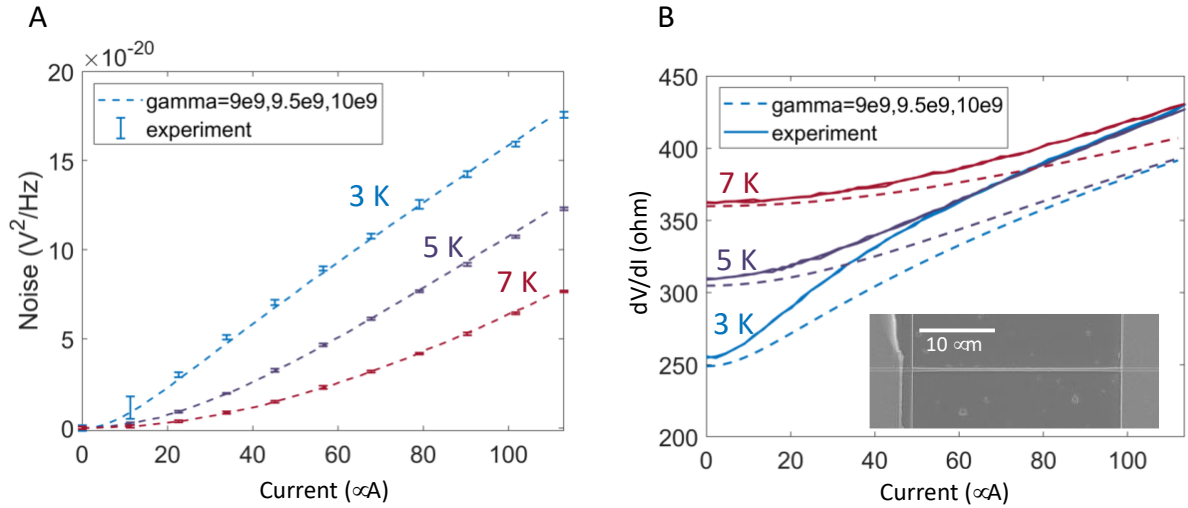

**Fig. S3. Inferring the electron-phonon coupling.** (A) Voltage noise as a function of bias current in a 30 μm long YbRh$_2$Si$_2$ wire, of comparable width to the nanowires used in the main experiment. The dashed curves are fits to the model of Sect. 5, where the fitted electron-phonon coupling parameters are $\Gamma = 9 \times 10^9$ K$^{-3}$m$^{-2}$, $9.5 \times 10^9$ K$^{-3}$m$^{-2}$, and $10 \times 10^9$ K$^{-3}$m$^{-2}$ at T=3 K, 5 K, and 7 K, respectively. (B): Using the temperature profiles computed from the model with those values of $\Gamma$, we compare the computed $\frac{dV}{dI}$ as a function of bias (dashed lines) with the measured data (solid lines), showing good qualitative agreement while implying some intrinsic non-Ohmic response in the material. The inferred $\Gamma$ values are of the same order as in gold, thus demonstrating that there is no greatly enhanced electron-phonon coupling in YbRh$_2$Si$_2$.

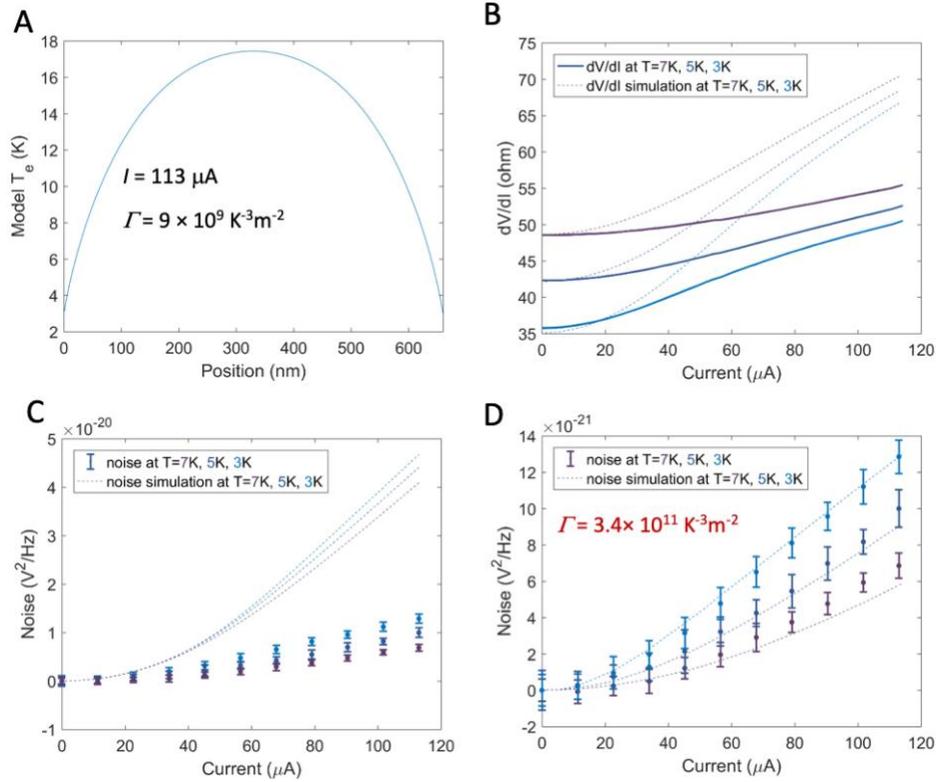

**Fig. S4. Simulations showing that the measured electron-phonon coupling cannot be the source of noise suppression**. (A) The modeled local electron temperature in the wire (calculated from Eq. (S2)) if we assume electronic heating and use the measured value of the electron-phonon coupling for YbRh$_2$Si$_2$ found from the experiments of Fig. S3 ($\Gamma = 9 \times 10^9$ K$^{-3}$m$^{-2}$ at the lowest temperature and largest current). (B) The calculated differential resistance vs. bias (dashed lines) expected from the bias-dependent temperature profiles as in (A), which disagree greatly with the measured values from device #1 (solid lines). (C) The calculated noise expected from the model (dashed lines), which disagree greatly with the measured values from device #1 (solid points). This shows that with the measured electron-phonon coupling (Fig. S3), in the usual quasiparticle/Fermi liquid treatment, there should be much greater noise than what is seen in the experiment. (D) Achieving the experimentally observed noise suppression in Yb$_2$Rh$_2$Si$_2$ through electron-phonon scattering would require an electron-phonon coupling of $3.4 \times 10^{11}$ K$^{-3}$m$^{-2}$, much larger than the measured value. Again, this shows that electron-phonon coupling cannot be responsible for the observed noise suppression.

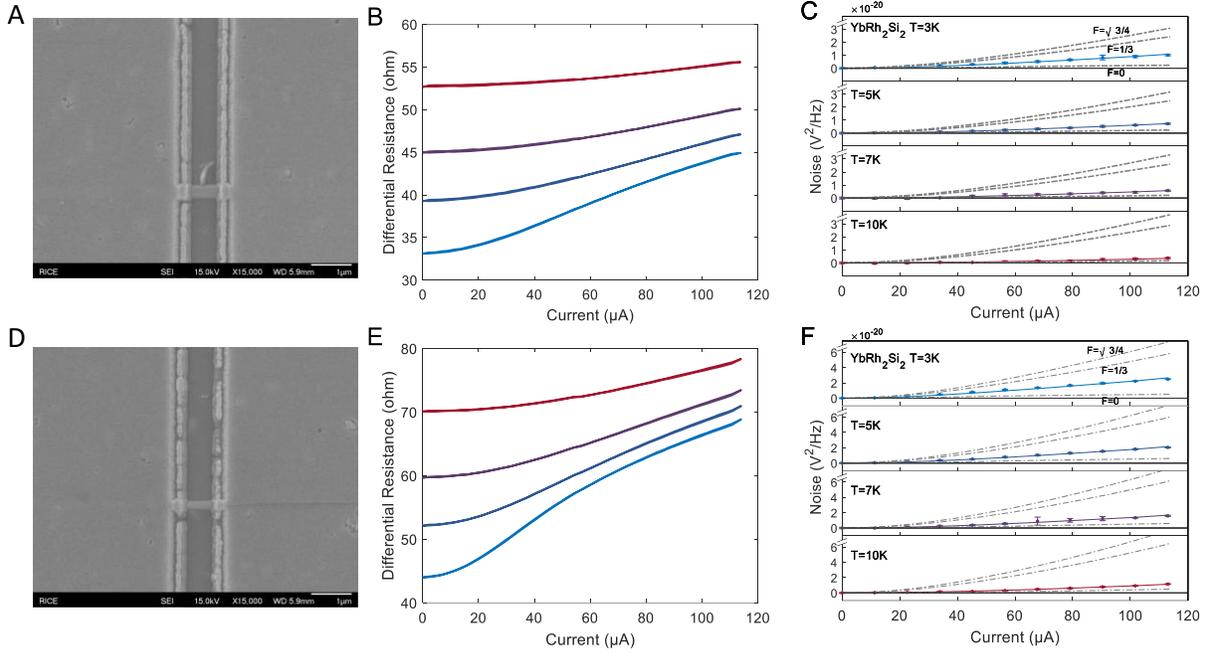

**Fig. S5. Additional data on two more YbRh$_2$Si$_2$ nanowire devices from the same film.** (A) SEM image of device #2. The nano wire in middle is connected by the two large gold pads on the two sides. (B) Differential resistance $dV/dI$ dependence on bias current at multiple temperatures. The curves from top to bottom are for 10 K, 7 K, 5 K, and 3 K respectively. (C) Noise vs. bias current at the same temperatures, with comparison dashed lines showing expectations for particular Fano factors. The measured noise remains far below theoretical expectations for a diffusive nanowire of a conventional Fermi liquid at all temperature. (D-F) SEM image of device #3; differential resistance dependence on bias current at 10 K, 7 K, 5 K, and 3 K; noise vs. bias current at the same temperatures. These results are all consistent with those for device #1 in the main text.

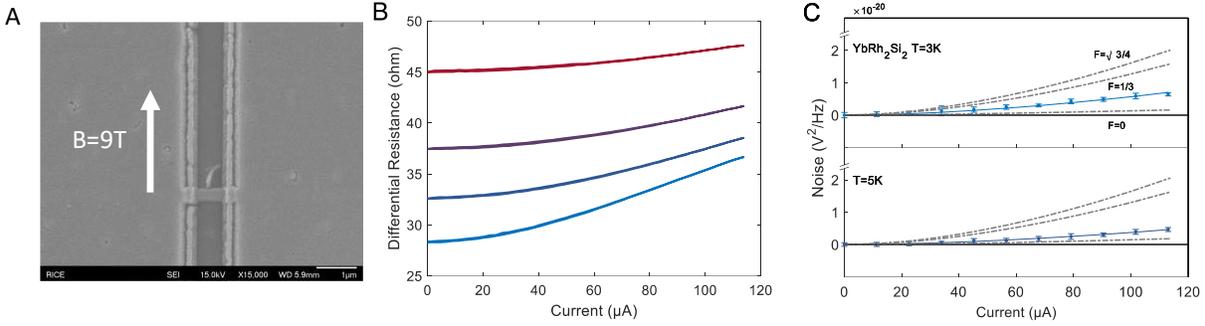

**Fig. S6. Additional data on YbRh$_2$Si$_2$ nanowire device #2 at high magnetic field.** (A) SEM image of the device indicating the direction of magnetic field. (B) Differential resistance $dV/dI$ dependence on bias current at multiple temperatures. The curves from top to bottom are for 10 K, 7 K, 5 K, and 3 K respectively. (C), (D) Noise vs. bias current at various temperatures 3 K and 5 K. The data are plotted in the same way as zero magnetic field, with comparison dashed lines showing expectations for particular Fano factors and error bars showing the experiment data. The noise intensity shows same dependence on bias and temperature as in the data in Fig. S5A, except for a small difference in overall value due to the slightly varied differential resistance.

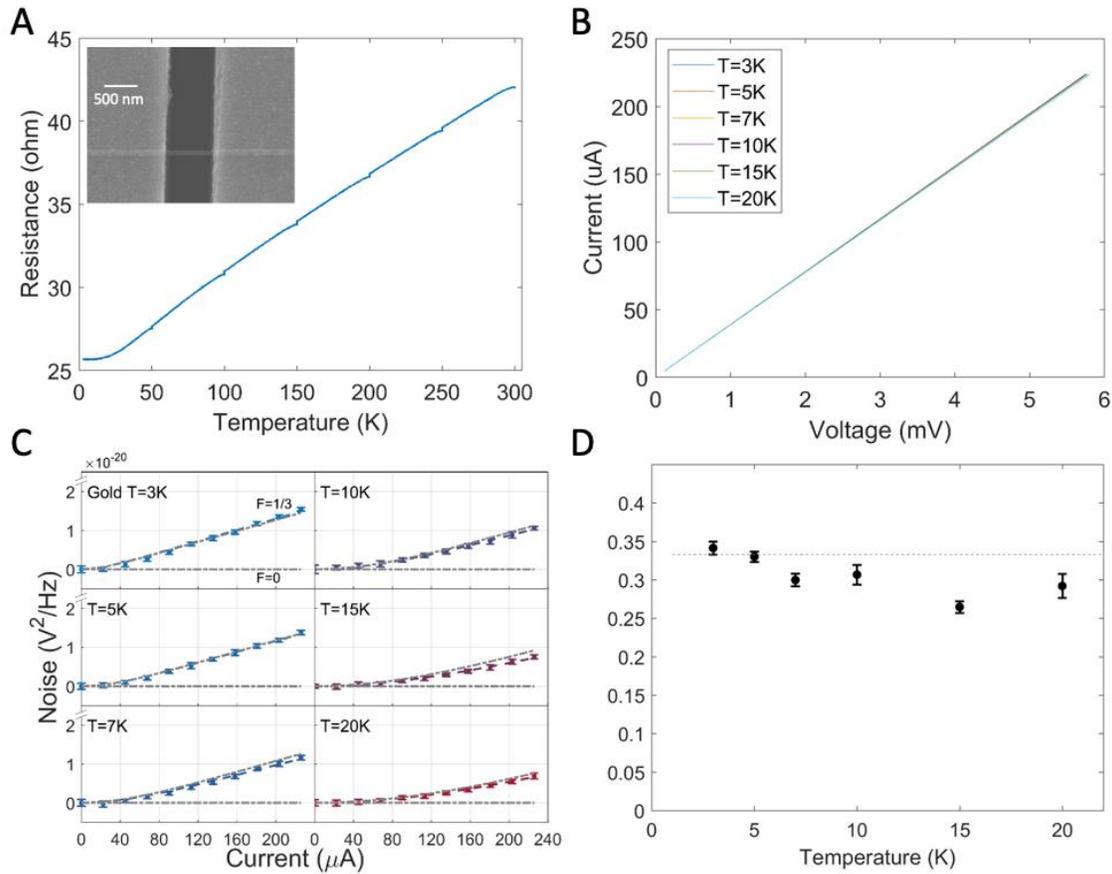

**Fig. S7. Additional data on a gold wire device.** (A) Resistance dependence on temperature of the Au wire (shown in an electron micrograph in the inset). The resistance decreases with decreasing temperature and saturates quadratically below 10 K. (B) Representative *IV* characteristics of the Au nanowire at temperatures from 3 K to 20 K. (C) Noise vs. bias current at various temperatures (3 K, 5 K, 7 K, 10 K, 15 K, 20 K) plotted as in Fig. 3B. (D) The Fano factor of the gold nanowire decreases slightly as temperature increase from 3 K to 20 K, likely due to the onset of electron-phonon scattering at the higher temperatures, and remains much larger than that of the $YbRh_2Si_2$ devices across the whole temperature range.

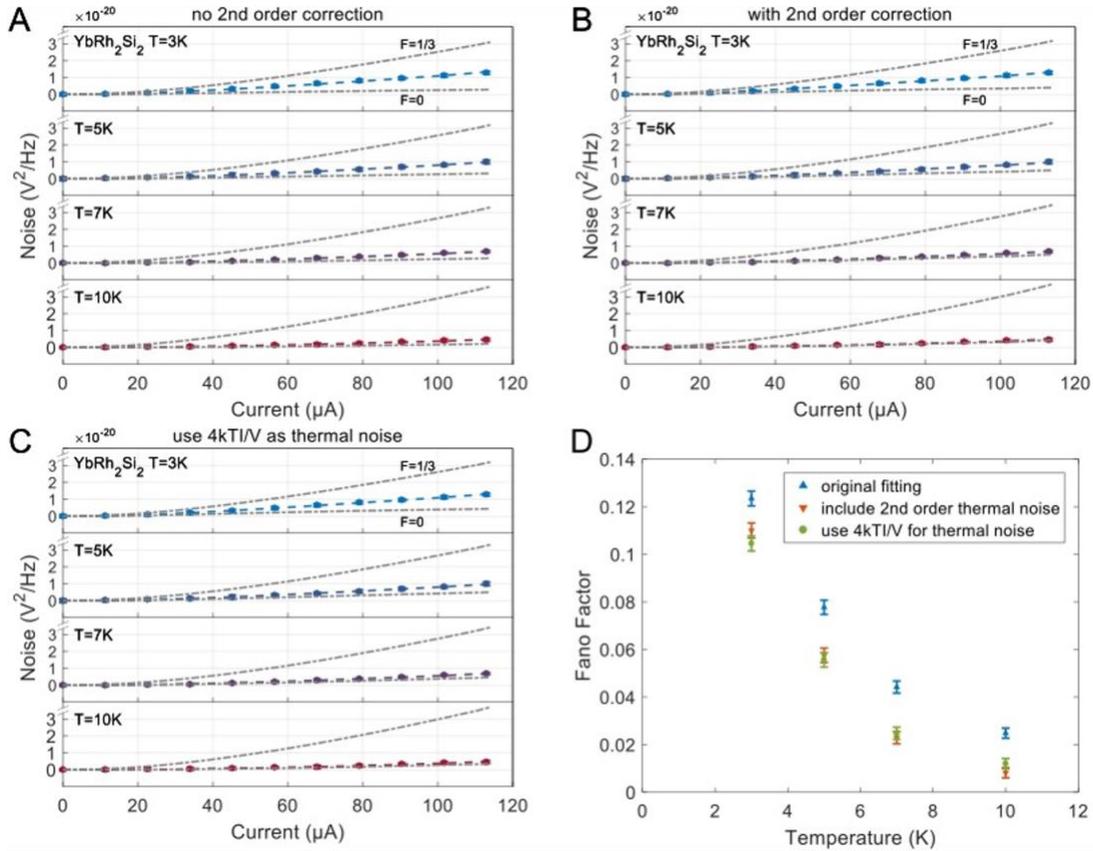

**Fig. S8**. Noise vs. bias current with two different analyses (curves shifted vertically for clarity) for device #1 from the main manuscript. The noise data are identical between the two panels and clearly fall far below the $F = \frac{1}{3}$ Fermi liquid expectations. (A) The theory curves in this panel including the fits to the data are using the expression from the paper, reproduced as Eq. (S3). (B) The theory curves in this panel including fits to the data are using an expression based on the second-order correction from M. S. Gupta, *Proc. IEEE* **70**, 788-804 (1982), shown in Eq. (S4). (C) The theory curves in this panel are based on the expression in Eq. (S6). (D) Comparison of fitted Fano factors from these three approaches. This shows that including a thermal noise correction that assumes an intrinsic non-Ohmic response *lowers* the inferred Fano factors by about 0.01 to 0.02.

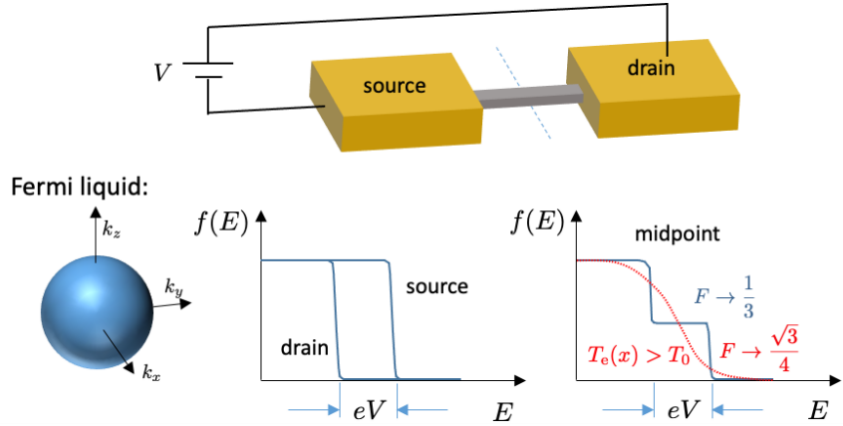

**Fig. S9**. **A qualitative picture of noise in the Fermi liquid case**. In the Fermi liquid, quasiparticles in the source and drain are FD distributed. In the absence of electron-electron interactions, the nonequilibrium quasiparticle distribution function in the diffusive wire has two steps and leads to the predicted $F = 1/3$ noise. In the strong electron-electron scattering limit, the nonequilibrium quasiparticle distribution function is FD-like with an elevated local temperature, leading to $F = \sqrt{3}/4$ noise. The situation in a non-Fermi liquid without quasiparticles, with an equilibrium distribution function very different from the FD case, remains to be addressed.

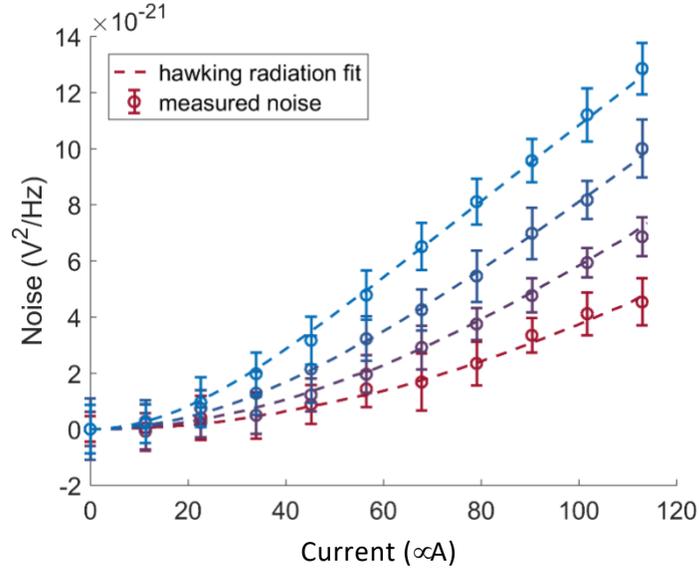

**Fig. S10. Comparison of measured noise with the example holographic treatment of noise in a quantum critical system.** The experimental data for device #1 from the main manuscript from 3 K to 10 K are plotted along with curves calculated using the expression mentioned in Sect. 12 and fit to the 3 K data.

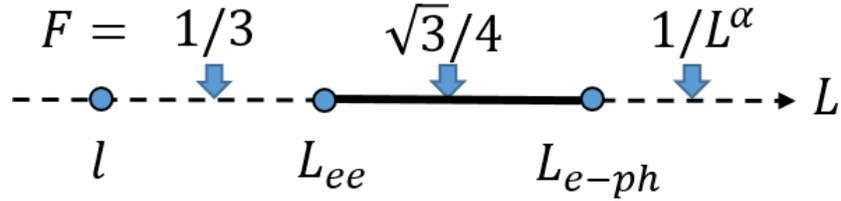

**Fig. S11. Behavior of the Fano factor ($F$) as a function of system size ($L$).** Within a Fermi liquid description, the Fano factor is sensitive to the system size relative to the scattering lengths resulting from electron-impurity scatterings ($l$), electron-electron scatterings ($L_{ee}$), and electron-phonon scatterings ($L_{ph}$). For a strongly correlated Fermi liquid metal, the regime of interest is marked by the solid line.